\documentclass[12pt]{article}
\usepackage{latexsym,amsmath,amsfonts,amssymb,eepic}
\usepackage{graphicx}
\newtheorem{theorem}{Theorem}

\newtheorem{lemma}{Lemma}

\newenvironment{proof}[1][Proof]{\textbf{#1.} }{\ \rule{0.5em}{0.5em}}
\begin{document}
\title{On the approximability of the vertex cover and related
 problems}
\author{Qiaoming Han
\thanks{Department of Management Science and
Engineering, Nanjing University, P.R.China. Email: qmhan@nju.edu.cn.
}  \and Abraham P. Punnen\thanks{Department of Mathematics, Simon
Fraser University, 14th Floor Central City Tower, 13450 102nd Ave.,
Surrey, BC V3T5X3, Canada. E-mail: apunnen@sfu.ca.} }
\date{}
\maketitle
\begin{abstract}
In this paper we show that the problem of identifying an edge
$(i,j)$ in a graph $G$ such that there exists an optimal vertex
cover $S$ of $G$ containing exactly one of the nodes $i$ and $j$ is
NP-hard. Such an edge is called a weak edge. We then develop  a
polynomial time approximation algorithm for the vertex cover problem
with performance guarantee $2-\frac{1}{1+\sigma}$, where $\sigma$ is
an upper bound on a measure related to a weak edge of a graph.
Further, we discuss a new relaxation of the vertex cover problem
which is used in our approximation algorithm to obtain smaller
values of $\sigma$. We also obtain linear programming
representations of the vertex cover problem for special graphs. Our
results provide new insights into the approximability of the vertex
cover problem - a long standing open problem.
\end{abstract}
{\bf Keywords:} vertex cover problem, approximation algorithm,
LP-relaxation, weak edge reduction, NP-complete problems
\par

\section{Introduction}
Let $G=(V, E)$ be an undirected graph on the vertex set
$V=\{1,2,\ldots ,n\}$. A {\it vertex cover} of $G$ is  a subset $S$
of $V$ such that each edge of $G$ has at least one endpoint in $S$.
The {\it vertex cover problem} (VCP)  is to compute a vertex cover
of smallest cardinality in $G$. The VCP is NP-hard on an arbitrary
graph but solvable in polynomial time on a bipartite graph. A vertex
cover $S$ is said to be $\gamma $-optimal if $|S|\leq \gamma |S^0|$ where $\gamma \ge 1$ and
$S^0$ is an optimal solution to the VCP.

\vskip 5pt

It is well known that a 2-optimal vertex cover of a graph can be
obtained in polynomial time by taking all the vertices of a maximal
(not necessarily maximum) matching in the graph or rounding up the
LP relaxation solution of an integer programming
formulation~\cite{nt75}. There has been considerable work (see e.g.
survey paper \cite{hoch}) on the problem over the past 30 years on
finding a polynomial-time approximation algorithm with an improved
performance guarantee. The current best known bound on the
performance ratio of a polynomial time approximation algorithm for
VCP is $2-\Theta(\frac{1}{\sqrt{\log n}})$~\cite{kara05}. It is also known that computing
a $\gamma$-optimal solution in polynomial time for VCP is NP-Hard
for any $1\leq \gamma \leq 10\sqrt{5}-21\simeq 1.36$~\cite{dinur}.
In fact, no polynomial-time $(2-\epsilon )$-approximation algorithm
is known for VCP for any constant $\epsilon >0$ and existence of
such an algorithm is one of the most outstanding open questions in
approximation algorithms for combinatorial optimization problems.
Under the assumption that the unique game
conjecture~\cite{harb,khot1,khot} is true, many researchers believe
that a polynomial time $(2-\epsilon)$-approximation algorithm with
constant $\epsilon >0$ is not possible for VCP. For recent works on
approximability of VCP, we refer
to~\cite{arora,bar85,charikar,dinur,halperin02,hpy07,hastad97,kara05,goemans98,monien85}.
Recently Asgeirsson and Stein~\cite{stein,stein1} reported extensive
experimental results using a heuristic algorithm which obtained no
worse than $\frac{3}{2}$-optimal solutions for all the test problems
they considered. Also, Han, Punnen and Ye~\cite{hpy07} proposed a
$(\frac{3}{2}+\xi)$-approximation algorithm for VCP, where $\xi$ is
an error parameter calculated by the algorithm and reported that no
example was known where $\xi\ne 0$.

\vskip 5pt
\vskip 5pt
A natural integer programming formulation of VCP can be described as follows:\\
\begin{equation}
\label{vc}(IP) \hskip 15pt
\begin{array}{lcl}
&\min & \sum_{i=1}^nx_i \\
&s.t. & x_i+x_j\ge 1, (i,j)\in E,\\
 &    & x_i\in \{0, 1\}, i=1,2,\cdots,n.
\end{array}
\end{equation}
Let $\bar{x}=(\bar{x}_1,\bar{x}_2,\ldots ,\bar{x}_n)$ be an optimal
solution  to (\ref{vc}). Then  $S_{IP}=\{i\ | \ \bar{x}_i=1\}$ is an
optimal vertex cover of the graph $G$. The linear programming (LP)
relaxation of the above integer program is
\begin{equation}
\label{lp}(LPR) \hskip 15pt
\begin{array}{lcl}
&\min & \sum_{i=1}^nx_i \\
&s.t. & x_i+x_j\ge 1, (i,j)\in E,\\
 &    &  x_i\ge 0, i=1,2,\cdots,n.
\end{array}
\end{equation}
It is well known that (e.g. \cite{nt74}) any optimal basic feasible
solution (BFS) $x^*=(x^*_1, x^*_2, \ldots, x^*_n)$ to the problem
LPR, satisfies $x_i^*\in\{0, \frac 1 2, 1\}$.  Let $S_{LP}=\{i \ | \
x_i^*=\frac 1 2 \mbox{ or } x_i^*=1\}$, then it is easy to see that
$S_{LP}$ is a $2$-approximate solution to the VCP on graph $G$.
Nemhauser and Trotter \cite{nt75} have further proved that there
exists an optimal integer solution to (\ref{vc}), which agrees with
$x^*$ in its integer components.

\vskip 5pt

An $(i, j)\in E$ is said to be a {\it weak edge} if there exists an
optimal vertex cover $V^0$ of $G$ such that $ |V^0\cap \{i,j\}|=1$.
Likewise, an $(i,j)\in E$ is said to be a \emph{strong edge} if
there exists an optimal vertex cover $V^0$ of $G$ such that  $
|V^0\cap \{i,j\}|=2$. An edge $(i,j)$ is \emph{uniformly strong} if
$ |V^0\cap \{i,j\}|=2$ for any optimal vertex cover $V^0.$ Note that
it is possible for an edge to be both strong and weak. Also $(i,j)$
is uniformly strong if and only if it is not a weak edge. In this
paper, we show that the problems of identifying a weak edge and
identifying a strong edge are NP-hard. We also present a polynomial
time $(2-\frac{1}{\sigma+1})$-approximation algorithm for VCP where
$\sigma$ is an appropriate graph theoretic measure (to be introduced
in Section 3). Thus for all graphs for which $\sigma$ bounded above
by a constant, we have a polynomial time
$(2-\epsilon)$-approximation algorithm for VCP. So far we could not
identify any class of graphs where $\sigma$ is anything but a
constant. We also give examples of graphs satisfying the property
that $\sigma=0$. However, establishing tight bounds on $\sigma$,
independent of graph structures remains an open question.  VCP is
trivial on a complete graph $K_n$ since any collection of $n-1$
nodes serves as an optimal solution. However, the LPR
gives an objective function value of $\frac{n}{2}$ only. We give a
stronger relaxation for VCP and complete linear programming
description of VCP on a complete graph, wheels, among others. This
relaxation can also be used to find reasonable expected guarantee
for $\sigma$.

\vskip 5pt

For any graph $G$, we sometimes use the notation $V(G)$ to represent
its vertex set and $E(G)$ to represent its edge set.



\section{Complexity of weak and strong edge problems}

The \emph{strong edge problem} can be stated as follows: ``Given a
graph, identify a strong edge of $G$ or declare that no such edge
exists.''

\begin{theorem}\label{th1}The strong edge problem is NP-hard.\end{theorem}
\begin{proof}
If $G$ is bipartite, then it does not contain a strong edge. If $G$
is not bipartite, then it must contain an odd cycle. For any odd
cycle $\omega$, any vertex cover must contain at least two adjacent
nodes of $\omega$ and hence $G$ must contain at least one strong
edge. If such an edge $(i,j)$ can be identified in polynomial time,
then after removing the nodes $i$ and $j$ from $G$ and applying the
algorithm on $G-\{i,j\}$ and repeating the process we eventually
reach a bipartite graph for which an optimal vertex cover $\hat{V}$
can be identified in polynomial time. Then $\hat{V}$ together with
the nodes removed sofar will form an optimal vertex cover of $G$.
Thus if the strong edge problem can be solved in polynomial time,
then the VCP can be solved in polynomial time.
\end{proof}

\vskip 5pt

The problem of identifying a weak edge is much more interesting. The
\emph{weak edge problem} can be stated as follows: ``Given a graph
$G$, identify a weak edge of $G$.'' It may be noted that unlike a
strong edge, a weak edge exists for all graphs. We will now show
that the weak edge problem is NP-hard. Before discussing the proof
of this, we need to introduce some notations and definitions.

\vskip 5pt

Let $x^*=(x_1^*,x_2^*,\ldots ,x_n^*)$ be an optimal BFS
of  LPR, the linear programming relaxation  of the VCP. Let $I_0 =
\{i ~:~ x^*_i=0\}$ and $I_1 = \{i ~:~ x_i^*=1\}$. The graph
$\bar{G}=G\setminus \{I_0\cup I_1\}$ is called the $\{0,1\}$-reduced
graph of $G$. The process of computing $\bar{G}$ from $G$ is called
a \emph{\{0,1\}-reduction}.

\vskip 5pt

\begin{lemma} \cite{nt75}\label{z1} If $R$ is a vertex cover of $\bar{G}$ then
$R\cup I_1$ is a vertex cover of $G$. If $R$ is optimal for
$\bar{G}$, then $R\cup I_1$ is an optimal vertex cover for $G$. If $R$ is an
$\gamma$-optimal vertex cover of $\bar{G}$, then $R\cup I_1$ is an
$\gamma$-optimal vertex cover of $G$ for any $\gamma \geq
1$.
\end{lemma}

Let $(i,j)$ be an edge of $G$. Define  $\Delta_{ij}=\{k\ |\ (i,
k)\in E(G)\mbox{ and } (j, k)\in E(G)\}$, $D_i=\{s\in V(G)\ |\
(i,s)\in E(G), s\not=j, s\not\in\Delta_{ij}\}$, and $D_j=\{t\in
V(G)\ |\ (j,t)\in E(G), t\not=i, t\not\in\Delta_{ij}\}$. Construct
the new graph $G^{(i,j)}$ from $G$ as follows. From graph $G,$
delete $\Delta_{ij}$ and all the incident edges, connect each vertex
$s\in D_i$ to each vertex $t\in D_j$ whenever such an edge is not
already present, and delete vertices $i$ and $j$ with all the
incident edges. The operation of constructing $G^{(i,j)}$ from $G$
is called an \emph{$(i,j)$-reduction}. When $(i,j)$ is selected as a
weak edge, then the corresponding $(i,j)$-reduction is called a
\emph{weak edge reduction}. The weak edge reduction is a modified
version of the active edge reduction operation introduced
in~\cite{hpy07}.
\begin{center}
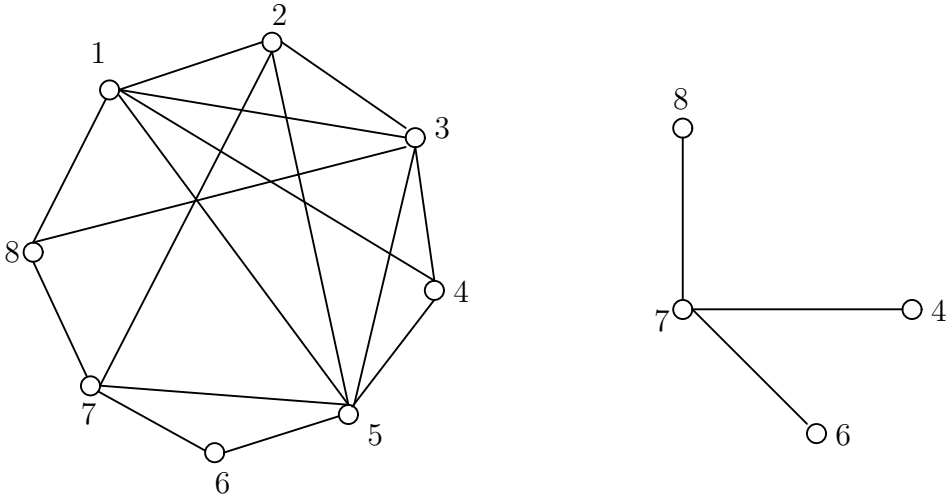
\begin{figure}[h]
\setlength{\unitlength}{0.254mm}
\begin{picture}(486,257)(30,-296)

        \allinethickness{0.254mm}\put(250,-190){\ellipse{10}{10}} 
        \allinethickness{0.254mm}\put(135,-275){\ellipse{10}{10}} 
        \allinethickness{0.254mm}\put(205,-255){\ellipse{10}{10}} 
        \allinethickness{0.254mm}\put(80,-85){\ellipse{10}{10}} 
        \allinethickness{0.254mm}\put(70,-240){\ellipse{10}{10}} 
        \allinethickness{0.254mm}\put(165,-60){\ellipse{10}{10}} 
        \allinethickness{0.254mm}\put(240,-110){\ellipse{10}{10}} 
        \allinethickness{0.254mm}\put(40,-170){\ellipse{10}{10}} 
        \allinethickness{0.254mm}\path(85,-85)(160,-60) 
        \allinethickness{0.254mm}\path(170,-60)(235,-105) 
        \allinethickness{0.254mm}\path(240,-115)(250,-185) 
        \allinethickness{0.254mm}\path(207,-251)(250,-195) 
        \allinethickness{0.254mm}\path(140,-275)(200,-256) 
        \allinethickness{0.254mm}\path(78,-90)(40,-165) 
        \allinethickness{0.254mm}\path(40,-175)(68,-235) 
        \allinethickness{0.254mm}\path(74,-243)(130,-274) 
        \allinethickness{0.254mm}\path(85,-85)(235,-110) 
        \allinethickness{0.254mm}\path(85,-85)(250,-185) 
        \allinethickness{0.254mm}\path(85,-88)(205,-250) 
        \allinethickness{0.254mm}\path(165,-65)(205,-250) 
        \allinethickness{0.254mm}\path(165,-65)(75,-240) 
        \allinethickness{0.254mm}\path(40,-165)(235,-115) 
        \allinethickness{0.254mm}\path(75,-240)(205,-250) 
        \allinethickness{0.254mm}\path(240,-115)(208,-250) 
        \put(70,-71){\shortstack{1}} 
        \put(165,-51){\shortstack{2}} 
        \put(250,-110){\shortstack{3}} 
        \put(260,-196){\shortstack{4}} 
        \put(215,-271){\shortstack{5}} 
        \put(135,-296){\shortstack{6}} 
        \put(65,-260){\shortstack{7}} 
        \put(25,-175){\shortstack{8}} 
        \allinethickness{0.254mm}\put(380,-105){\ellipse{10}{10}} 
        \allinethickness{0.254mm}\put(450,-265){\ellipse{10}{10}} 
        \allinethickness{0.254mm}\put(500,-200){\ellipse{10}{10}} 
        \allinethickness{0.254mm}\put(380,-200){\ellipse{10}{10}} 
        \allinethickness{0.254mm}\path(495,-200)(385,-200) 
        \allinethickness{0.254mm}\path(385,-200)(445,-260) 
        \allinethickness{0.254mm}\path(380,-110)(380,-195) 
        \put(375,-95){\shortstack{8}} 
        \put(365,-211){\shortstack{7}} 
        \put(510,-206){\shortstack{4}} 
        \put(460,-271){\shortstack{6}} 
\end{picture}

\caption{Weak edge reduction using edge $(1, 2)$. $D_1=\{8, 4\}$,
$D_2=\{7\}$ and $\Delta_{12}=\{3, 5\}$. The graph $G^{(1,2)}$ is on
the right.}
\end{figure}
\end{center}
\begin{lemma}\label{z2}Let $(i,j)$ be a weak edge of $G$, $R\subseteq V(G^{(i,j)})$
and \begin{equation*}R^* =\begin{cases} R\cup\Delta_{ij}\cup \{j\}, & \text{ if } D_i\subseteq R;\\
R\cup\Delta_{ij}\cup \{i\}, & \text{ otherwise},\end{cases}
\end{equation*} \begin{enumerate}
\item If $R$ is a vertex cover of $G^{(i,j)}$, then $R^*$ is a vertex
cover of $G$. \item If $R$ is an optimal vertex cover of $G^{(i,j)}$,
then $R^*$ is an optimal vertex cover of $G$. \item If $R$ is a
$\gamma$-optimal vertex cover in $G^{(i,j)}$, then $R^*$ is a
$\gamma$-optimal vertex cover in $G$ for any $\gamma \geq
1$.\end{enumerate}
\end{lemma}
\begin{proof}   If
$D_i\subseteq R$ then all arcs in $G$ incident on $i,$ except
possibly $(i,j),$ is covered by $R$. Then $R^*=R\cup\Delta_{ij}\cup
\{j\}$ covers all arcs incident on $j$, including $(i,j)$ and hence
$R^*$ is a vertex cover in $G$. If at least one vertex of $D_i$ is
not in $R$, then all vertices in $D_j$ must be in $R$ by
construction of $G^{(i,j)}$. Thus $R^*=R\cup\Delta_{ij}\cup\{i\}$
must be a vertex cover of $G$.

\vskip 3pt

 Suppose $R$ is an optimal vertex
cover of $G^{(i,j)}$. Since $(i,j)$ is a weak edge, there exists an
optimal vertex cover, say $V^0$, of $G$ containing exactly one of
the nodes $i$ or $j$. Without loss of generality, let this node be
$i$. For each node $k\in \Delta_{ij}$, $(i,j,k)$ is a 3-cycle in $G$
and hence $k\in V^0$ for all $k\in \Delta_{ij}$. Let
$V^1=V^0-(\{i\}\cup \Delta_{ij})$, which is a vertex cover of $G^{(i,j)}$. Then $|R|=|V^1|$ for otherwise if
$|R| < |V^1|$ we have $|R^*|<|V^0|$, a contradiction. Thus
$|R^*|=|V^0|$ establishing optimality of $R^*$.

\vskip 3pt

Suppose $R$ is an $\gamma$-optimal vertex cover of $G^{(i,j)}$ and
let $V^{(i,j)}$ be an optimal vertex cover in $G^{(i,j)}$. Thus
\begin{equation}\label{peq1} |R|\leq \gamma |V^{(i,j)}| \mbox{ where }\gamma \geq 1.\end{equation} Let
$V^0$ be an optimal vertex cover in $G$. Without loss of generality
assume $i\in V^0$ and since $(i,j)$ is weak, $j\notin V^0.$ Let
$V^1=V^0-(\{i\}\cup \Delta_{ij}).$ Then $|V^1|=|V^{(i,j)}|.$ Thus we
from (\ref{peq1}), $|R|\leq \gamma |V^1|.$ Thus $$|R^*|\leq
\gamma |V^1|+|\Delta_{ij}|+1 \leq \gamma (|V^1|+|\Delta_{ij}|+1)
\leq \gamma |V^0|.$$ Thus $R^*$ is $\gamma$-optimal in $G.$
\end{proof}

\vskip 5pt

Suppose that  an oracle, say WEAK($G,i,j$), is available which with
input $G$ outputs two nodes $i$ and $j$ such that $(i,j)$ is a weak
edge of $G$. It may be noted that WEAK($G,i,j$) do not tell us which
node amongst $i$ and $j$ is in an optimal vertex cover. It simply
identifies the weak edge $(i,j)$. Using the oracle WEAK($G,i,j$), we
 develop an algorithm, called \emph{weak edge reduction
algorithm} or WER-algorithm to compute an optimal vertex cover of
$G$.

\vskip 5pt

The basic idea of the scheme is very simple. We apply $\{0,1\}$ and
weak edge reductions repeatedly until a null graph is reached, in
which case the algorithm goes to a backtracking step. We record the
vertices of the weak edge identified in each weak edge reduction
step but do not determine which one to be included in the output
vertex cover. In the backtrack step, taking guidance from lemma 2,
we choose exactly one of these two vertices to form part of the
vertex cover we construct. In this step, the algorithm computes a
vertex cover for $G$ using all vertices in $\Delta_{ij}$ removed in
the weak edge reduction steps, vertices with value 1 removed in the
$\{0,1\}$ reduction steps, and the selected vertices in the
backtrack step from the vertices corresponding to the weak edges
recorded during the weak edge reduction steps. A formal description
of the WER-algorithm is given below.

\begin{itemize}
\item[~] {\large\bf The WER-Algorithm}
\item[Step 1:~] \{* \textsf{Initialize} *\}\
$ k=1, G_k=G$.
\item[Step 2:~] \{* \textsf{Reduction operations}
*\}\ $\Delta_k=\emptyset$, $I_{k,1}=\emptyset,$ $(i_k,j_k)=\emptyset$.
\begin{enumerate}
\item \{* \textsf{\{0,1\}-reduction} *\} Solve the LP relaxation problem LPR of VCP on the graph $G_k$. Let
$x^k=\{x_i^k : i\in V(G_k)\}$ be the resulting optimal BFS,
$I_{k,0}=\{i\ | \ x_i^k=0\}, \
I_{k,1}=\{ i\ | \ x_i^k=1\}$, and $I_k=I_{k,0}\cup I_{k,1}$.\\
\textbf{If} $V(G_k)\setminus I_{k}=\emptyset$ \textbf{goto} Step 3
\textbf{else}  $G_k=G_k\setminus I_k$ \textbf{endif}

\item \{* \textsf{weak edge reduction} *\}
 Call WEAK($G_k,i,j$) to identify the weak
edge $(i,j)$. Let $G_{k+1}=G_k^{(i,j)}$, where $G_k^{(i,j)}$ is the
graph obtained from $G_k$ using  the weak edge reduction operation.
Compute $\Delta_{ij}$ for $G_k$ as defined in the weak edge
reduction.
Let $\Delta_k=\Delta_{ij}, i_k=i, j_k=j$. \\
\textbf{If} $G_{k+1}\not=\emptyset$ \textbf{then} $k=k+1$
\textbf{goto} beginning of Step 2 \textbf{endif}
\end{enumerate}
\item[Step 3:~] L=k+1, $S_L=\emptyset$.
\item[Step 4:~] \{* \textsf{Backtracking to construct a solution} *\}\\
Let $S_{L-1}=S_{L}\cup I_{L-1,1}$,\\
\textbf{If} $(i_{L-1},j_{L-1})\neq \emptyset$ \textbf{then}
 $S_{L-1}=S_{L-1}\cup \Delta_{L-1}\cup R^*$, where
\begin{equation*}R^* =
\begin{cases} j_{L-1}, & \text{ if } D_{i_{L-1}}\subseteq S_L;\\
i_{L-1}, & \text{ otherwise},\end{cases}
\end{equation*}
$\mbox{ ~ }$ and $D_{i_{L-1}}=\{s  : (i_{L-1},s)\in
G_{L-1},s\ne
j_{L-1}, s\not\in\Delta_{L-1}\}$ \textbf{endif}\\
$ L = L-1$,\\ \textbf{If} $L\neq 1$ \textbf{then} \textbf{goto}
beginning of step 4 \textbf{else} output $S_1$ and STOP
\textbf{endif}
 \end{itemize}

\vskip 5pt

Using Lemma 1 and Lemma 2, it can be verified that the output $S_1$
of the WER-algorithm is an optimal vertex cover of $G$. It is easy
to verify that the complexity of the algorithm is polynomial
whenever the complexity of WEAK($G,i,j$) is polynomial. Since VCP is
NP-hard we established the following theorem:

\begin{theorem}\label{th3}The weak edge problem is
NP-hard.\end{theorem}

\section{An approximation algorithm for VCP}

Let VCP($i,j$) be the \emph{restricted vertex cover problem} where
feasible solutions are vertex covers of $G$ using exactly one of the
vertices from the set $\{i,j\}$ and looking for the smallest vertex
cover satisfying this property. More precisely, VCP($i,j$) tries to
identify a vertex cover $V^*$ of $G$ with smallest cardinality such
that $|V^*\cap \{i,j\}|=1.$ Let $\delta$ and $\bar{\delta}(i,j)$ be
the optimal objective function values of VCP and VCP($i,j$)
respectively. If $(i,j)$ is indeed a weak edge of $G$, then $\delta
= \bar{\delta}(i,j)$. Otherwise,
\begin{equation}
\label{cj2} \bar{\delta}(i,j)= \delta+\sigma(i,j),
\end{equation}
where $\sigma(i,j)$ is a non-negative integer. Further, using
arguments similar to the proof of Lemma 2 it can be shown that
\begin{equation}\label{peq2}
\zeta_{ij}+\Delta_{ij}+1=\bar{\delta}(i,j)=\delta+\sigma(i,j).
\end{equation} where $\zeta_{ij}$ is the optimal objective function
value VCP on $G^{(i,j)}$.

Consider the optimization problem
\begin{tabbing}
XXXXXXXXXXXXX\=XXXXXXXXXXXX\=XXXXXXXXXXX\=XXX \kill

\>WEAK-OPT: \>Minimize $\sigma(i,j)$\\
\>\> Subject to $(i,j)\in E(G)$
\end{tabbing}

WEAK-OPT is precisely the weak edge problem in the optimization form
and its optimal objective function value is always zero. However
this problem is NP-hard by Theorem~\ref{th3}. We now show that an
upper bound $\sigma$ on the optimal objective function value of
WEAK-OPT and a solution $(i,j)$ with $\sigma(i,j)\leq \sigma$ can be
used to obtain a $(2-\frac{1}{1+\sigma})$-approximation algorithm
for VCP. Let ALMOST-WEAK($G,i,j$) be an oracle which with input $G$
computes an approximate solution $(i,j)$ to WEAK-OPT such that
$\sigma(i,j)\leq \sigma$ for some $\sigma$. Consider the
WER-algorithm with WEAK($G,i,j$) replaced by ALMOST-WEAK($G,i,j$).
We call this the AWER-algorithm.

Let $G_k,~k=1,2,\ldots t$ be the sequence of graphs generated in
Step 2(2) of the AWER-algorithm and $(i_k,j_k)$ be the approximate
solution to WEAK-OPT on $G_k,~k=1,2,\ldots ,t$ identified by
ALMOST-WEAK($G_k,i_k,j_k$).

\begin{theorem}
\label{conclusion2} The AWER-algorithm identifies a vertex cover
$S_1$ such that $|S_1|\leq (2-\frac{1}{1+\sigma})|S^*|$ where $S^*$
is an optimal solution to the VCP. Further, the complexity of the
the algorithm is $O(n(\phi(n)+\psi(n))$ where $n=|V(G)|$, $\phi(n)$
is the complexity of LPR and $\psi(n)$ is the complexity of
ALMOST-WEAK($G,i,j$).
\end{theorem}
\begin{proof}Without loss of generality, we assume that the LPR solution
$x^1=(x_1^1,x^1_2,\ldots ,x^1_n)$ generated when Step 2(1) is
executed for the first time satisfies $x^1_i=\frac{1}{2}$ for all
$i$. If this is not true, then we could replace $G$ by a new graph
$\bar{G}=G \setminus \{I_{1,1}\cup I_{1,0}\}$ and by Lemma 1, if
$\bar{S}$ is a $\gamma$-optimal solution for VCP on $\bar{G}$ then
$\bar{S}\cup I_{1, 1}$ is a $\gamma$-optimal solution on $G$ for any
$ \gamma \ge 1$. Thus, under this assumption we have
\begin{equation}\label{x1}n\leq
2|S^*|.\end{equation}
 Let $t$ be the total number of iterations of
Step 2 (2). For simplicity of notation, we denote
$\sigma_k=\sigma(i_k,j_k)$ and
$\bar{\delta}_k=\bar{\delta}(i_k,j_k)$. Note that $\delta_k$ and
$\bar{\delta_k}$ are optimal objective function values of VCP and
VCP($i_k,j_k$), respectively, on the graph $G_k$. In view of
equations (\ref{cj2}) and (\ref{peq2}) we have,
\begin{equation}\label{x2}
\bar{\delta}_k=\delta_k+\sigma_k,~~~~~k=1,2,\ldots ,t
\end{equation} and
\begin{equation}\label{x3}
\delta_{k+1}+|\Delta_{i_k,j_k}|+|I_{k,1}|+1=\bar{\delta}_k,
~~~~~k=1,2,\ldots ,t.
\end{equation} From  (\ref{x2}) and (\ref{x3}) we have
\begin{equation}\label{x5}
\delta_{k+1}-\delta_k=\sigma_k-|\Delta_{i_k,j_k}|-|I_{k,1}|-1,~~~k=1,2,\ldots
,t.\end{equation} Adding equations in (\ref{x5}) for $k=1,2,\ldots
,t$ and using the fact that $\delta_{t+1}=|I_{t+1,1}|$, we have,
\begin{equation}\label{x4}
|S_1|=|S^*|+\sum_{k=1}^t\sigma_k,
\end{equation}
where $|S^*|=\delta_1$, and by construction,
\begin{equation}\label{x7}
|S_1|=\Sigma_{k=1}^{t+1}I_{k,1}+\Sigma_{k=1}^t\Delta_k+t.
\end{equation}
 But,
\begin{equation}\label{x6}
|V(G)|=\Sigma_{k=1}^{t+1}I_{k}+\Sigma_{k=1}^t\Delta_k+2t.
\end{equation}
From (\ref{x4}), (\ref{x7}) and (\ref{x6}), we have
\begin{equation}\label{e2}
t=\frac{|V(G)|-\Sigma_{k=1}^{t+1}I_{k}-\Sigma_{k=1}^t\Delta_k}{2}
\le \frac{|V(G)|-|S^*|-\Sigma_{k=1}^t(\sigma_k-1)}{2}.
\end{equation}
From inequalities~(\ref{x1}) and (\ref{e2}), we have
\begin{equation*}
t\le \frac{|S^*|-t(\bar{\sigma}-1)}{2},
\end{equation*}
where $\bar{\sigma}=\frac{\Sigma_{k=1}^t\sigma_k}{t}$.  Then we have
\[
t\le\frac{|S^*|}{\bar{\sigma}+1}.
\]
Thus,
\[
\frac{|S_1|}{|S^*|} =
\frac{|S^*|+\Sigma_{k=1}^t\sigma_k}{|S^*|}=\frac{|S^*|+t\bar{\sigma}}{|S^*|}\le
1+\frac{\bar{\sigma}}{\bar{\sigma}+1}\le
1+\frac{\sigma}{\sigma+1}=2-\frac{1}{1+\sigma}.
\] The complexity of the algorithm can easily be verified.
\end{proof}

\vskip 5pt

The performance bound established in Theorem~\ref{conclusion2} is
useful only if we can find an efficient way to implement our
black-box oracle ALMOST-WEAK($G,i,j$) that identifies a reasonable
$(i,j)$ in each iteration. If ALMOST-WEAK($G,i,j)$ simply generates
a random edge, then $\sigma(i,j)$ could be as large as $O(n)$ as
given in the example of Figure~2.

\begin{center}
\begin{figure}[h]
    ~~~~~~~~~~~~~~~~~~~~~~~~~~~~~~~~~~\includegraphics[width=4.5cm, height=4.5cm]{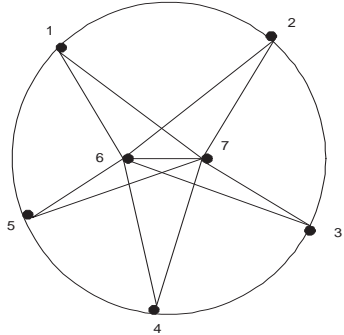}
  \caption{A 3D-wheel with central axis (6, 7).}\label{fig2}
\end{figure}
\end{center}

For a 3D-wheel on $n$ nodes with central axis  $(n-1, n)$,
$\sigma(n-1, n)=\lfloor\{\frac{n}{2}\}-2\rfloor$. However, when
$(i,j)$ is chosen as any other edge, $\sigma(i,j)=0$.  A trivial
upper bound on $\sigma$ is $\frac{n}{2}$ for any graph on $n$ nodes.
Let us now explore the possibilities of improving this trivial
bound.

\vskip 8pt

Any vertex cover must contain at least $s+1$ vertices of an odd
cycle of length  $2s+1$. This motivates  the following
\emph{extended linear programming relaxation} (ELP)  of the VCP,
 studied in~\cite{arora,hpy07}.

\begin{equation}
\label{elp}(ELP) \hskip 10pt
\begin{array}{lcl}
&\min & {\displaystyle \sum_{i=1}^nx_i} \\
&s.t. & x_i+x_j\ge 1, (i,j)\in E,\\
 &    & {\displaystyle\sum_{i\in \omega_k}x_i\ge s_k+1, \omega_k\in \Omega,}\\
  &   &  x_i\ge 0, i=1,2,\ldots,n,
\end{array}
\end{equation}
where $\Omega$ denotes the set of all odd-cycles of  $G$ and
$\omega_k\in\Omega$ contains $2s_k+1$ vertices for some integer
$s_k$. Note that although there may be an exponential number of
odd-cycles in~$G$, since the odd cycle inequalities has a
polynomial-time separation scheme, ELP is polynomially solvable.
Further, it is possible to compute an optimal BFS of ELP in polynomial time.

\vskip 5pt

Let $x^0$ be an optimal basic feasible solution of ELP. An edge $(r,
s)\in E$ is said to be an {\it active edge} with respect to $x^0$ if
$x^0_i+x^0_j=1$. There may or may not exist an active edge
corresponding to an optimal BFS of the ELP as shown in~\cite{hpy07}.
For any arc $(r,s)$, consider the \emph{restricted ELP}
(RELP($r,s$)) as follows:

\begin{equation}
\label{elp}(RELP(r,s)) \hskip 15pt
\begin{array}{lcl}
&\min & {\displaystyle\sum_{i=1}^nx_i }\\
&s.t. & x_i+x_j\ge 1, (i,j)\in E\backslash \{(r,s)\},\\
&  &  x_r+x_s = 1,\\
 &    & {\displaystyle\sum_{i\in \omega_k}x_i\ge s_k+1, \omega_k\in \Omega,}\\
  &   &  x_i\ge 0, i=1,2,\ldots,n,
\end{array}
\end{equation}
Let $Z(r,s)$ be the optimal objective function value of RELP($r,s$).
Choose $(p,q)\in E(G)$ such that
$$\quad \quad \quad Z(p,q) = \min\{Z(i,j)~:~(i,j)\in E(G)\}.$$
An optimal solution to RELP($p,q$) is called \emph{a RELP solution}.
It may be noted that if an optimal solution $x^*$ of the ELP
contains an active edge, then $x^*$ is also an RELP solution.
Further $Z(p,q)$ is a lower bound on the optimal objective function
value of VCP.

The VCP on a complete graph is trivial since any collection of
$(n-1)$ nodes form an optimal vertex cover. However, for a complete
graph, LPR yields an optimal objective function value of
$\frac{n}{2}$ only and ELP yields an optimal objective function
value of $\frac{2n}{3}$. Interestingly, the optimal objective
function value of RELP on a complete graph is  $n-1$, and the RELP
solution is indeed an optimal vertex cover on a complete graph. A
stronger version of this observation is proved in the following
theorem.

\begin{theorem}For any $(i,j)\in E(G),$ an optimal BFS of the linear program
RELP($i,j$) gives an optimal vertex cover of $G$  whenever $G$ is a
complete graph or a wheel.
\end{theorem}
\begin{proof}Suppose $G$ is a complete graph with $V(G)=\{1,2,\ldots ,n\}.$
 Without loss of generality assume $(i,j)=(1,2)$. Thus
\begin{equation}\label{t2}x^0_1+x^0_2=1.\end{equation}
Let $x^0=(x_1^0,x_2^0,\ldots ,x_n^0)$
be an optimal basic solution of RELP($i,j$). Since $x_1^0+x_2^0=1,$
by odd cycle inequalities, we have
\begin{equation}\label{q1}x_1^0+x_2^0+x_k^0 = 2,~~~k=2,3,\ldots ,n.
\end{equation} Hence $x_k^0=1$ for $k=3,4,\ldots ,n$ yielding $Z(i,j)=n-1$.
Now we have to establish that $x^0_1$ and $x^0_2$ cannot be
fractional. If $x^0_1+x^0_r+x^0_s=2$ for any $\{r,s\}\ne\{1,2\}$
then $x^0_1=0$ yielding $x^0_2=1$. Similarly if
$x^0_2+x^0_r+x^0_s=2$ for any $\{r,s\}\ne\{1,2\}$ then $x^0_2=0$
yielding $x^0_1=1$. If $x^0_r+x^0_s+x^0_t > 2$ for all 3-cycles
other than those in (\ref{q1}) it can be shown that there must exist
an edge inequality, other than (\ref{t2}), satisfied as an equality.
Such an equality must be of the form $x^0_1+x^0_r=1$ or
$x^0_2+x^0_r=1$ for $r\in \{3,4,\ldots ,n\}$ and hence $x^0_1$ and
$x^0_2$ can take only values zero or one. Both $\{2,3,\ldots ,n\}$
and $\{1,3,\ldots ,n\}$ are optimal vertex covers for $G$. The proof
for the case of a wheel can  be obtained using similar analysis and
we skip the details.
\end{proof}

\vskip 5pt

It may be noted that an optimal BFS of RELP($i,j$) gives an optimal
vertex cover on a 3D-wheel (Figure~2) when $(i, j)$ is not the
central axis.

\vskip 8pt

Extending the notion of an active edge corresponding to an ELP
solution~\cite{hpy07}, an edge $(i, j)\in E$ is said to be an {\it
active edge} with respect to an RELP solution $x^0$ if
$x^0_i+x^0_j=1$. Unlike ELP, an RELP solution always contains an
active edge. In AWER-algorithm, the output of ALMOST-WEAK($G,i,j$)
can be selected as an active edge with respect to a RELP solution on
$G$.

\vskip 5pt

We believe that the value of $\sigma(i,j)$, i.e. the absolute
difference between the optimal objective function value of VCP and
the optimal objective function value of VCP($i,j$), when $(i,j)$ is
an active edge corresponding to an RELP solution is a constant with
very high probability, if not with probability one. If it is a
constant, then our algorithm resolves the long standing question on
the existence of a polynomial time $2-\epsilon$ approximation
algorithm for VCP for constant $\epsilon > 0.$ It is an open
question to obtain a tight bound, deterministic or probabilistic, on
this interesting graph theoretic measure. Nevertheless, our results
provide new insight into the approximability of the vertex cover
problem.

\section{Conclusion}
In this paper, we proved that weak edge and strong edge problems are
NP-hard. We also presented a polynomial time
$(2-\frac{1}{\sigma+1})$-approximation algorithm for VCP where
$\sigma$ is a well defined graph theoretic measure. Obtaining tight
upper bounds on $\sigma,$ deterministic or probabilistic,  is an
open question. We also provide simple linear programming
representation of VCP on a complete graph, wheel and 3D-wheel, among
other graphs.\\

\noindent {\bf Acknowledgements:} Research of Qiaoming Han was
supported in part by Chinese NNSF grant, and the Return-from-Abroad
foundation of MOE China. Research of Abraham P. Punnen is partially
supported by an NSERC discovery grant.
{ 


}

\begin{thebibliography}{999}
\bibitem{arora} S. Arora, B. Bollob$\acute{a}$s and L.
Lov$\acute{a}$sz, Proving integrality gaps without knowing the
linear program, {\it Proc. IEEE FOCS} (2002), 313-322.

\bibitem{stein} E. Asgeirsson and C. Stein, Vertex cover approximations on random graphs,
Lecture notes in computer science, Volume 4525 (2007) 285-296
Springer Verlag.

\bibitem{stein1} E. Asgeirsson and C. Stein, Vertex cover
approximations: Experiments and observations, WEA, 545-557 (2005).

\bibitem{bar85} R. Bar-Yehuda and S. Even, A local-ratio theorem for approximating the weighted vertex cover problem,
 {\it Annals of Discrete Mathematics}, 25(1985), 27-45.

\bibitem{charikar} Moses Charikar, On semidefinite programming
relaxations for graph coloring and vertex cover, {\it Proc. 13th
SODA} (2002), 616-620.

\bibitem{dinur} I. Dinur and S. Safra, The importance of being
biased, {\it Proc. 34th ACM Symposium on Theory of Computing},
33-42, 2002.

\bibitem{halperin02} Eran Halperin, Improved approximation algorithms for the vertex cover problem
in graphs and hypergraphs, {\it SIAM J. Comput.}, 31(2002),
1608-1623.

\bibitem{hpy07} Q. Han, A.P. Punnen and Y. Ye, A polynomial time $\frac 3 2$-approximation algorithm for the vertex cover problem on a class of graphs, Working paper, 2007.

\bibitem{harb} B. Harb, The unique games conjecture and some of its
implications on inapproximability. Manuscript, May 2005.

\bibitem{hastad97} J. H{\aa}stad, Some optimal inapproximability results,
{\it JACM} 48(2001), 798-859.



\bibitem{hoch} D. S. Hochbaum, Approximating covering and packing problems: set cover, independent set,
and related problems, in {\it Approximation Algorithms for NP-Hard
Problems}, 94-143, edited by D. S. Hochbaum, PWS Publishing Company,
1997.

\bibitem{kara05} G. Karakostas, A better approximation ratio for
the vertex cover problem, L. Caires et al. (Eds): ICALP 2005, LNCS
3580, 1043-1050.

\bibitem{khot1} S. Khot, On the power of unique 2-Prover 1-Round
games. In proceedings of 34th ACM symposium on Theory of Computing
(STOC) 767-775, 2002.

\bibitem{khot} S. Khot and O. Regev, Vertex cover might be hard to
approximate to within $2-\epsilon$. Manuscript.

\bibitem{goemans98} J. Kleinberg and M. Goemans, The Lov$\acute{a}$sz theta function and
a semidefinite programming relaxation of vertex cover, {\it SIAM J.
Discrete Math.}, 11(1998), 196-204.

\bibitem{monien85} B. Monien and E. Speckenmeyer, Ramsey numbers and an approximation algorithm for
the vertex cover problem, {\it Acta Informatica}, 22(1985), 115-123.

\bibitem{nt74} G. L. Nemhauser and L. E. Trotter, Jr. , Properties of vertex
packing and independence system polyhedra, {\it Mathematical
Programming}, 6(1974), 48-61.

\bibitem{nt75} G. L. Nemhauser and L. E. Trotter, Jr. , Vertex packings: Structural properties
and algorithms. {\it Mathematical Programming}, 8(1975), 232-248.
\end{thebibliography}
\end{document}